\documentclass[aps,prl,twocolumn,groupedaddress,showpacs]{revtex4-1}

\usepackage{graphicx}
\usepackage{braket}
\usepackage{amsmath}
\usepackage{textcomp}
\usepackage{color}
\usepackage[normalem]{ulem}

\begin{document}
\bibliographystyle{unsrt}

\title{Quantum computation with universal error mitigation on superconducting quantum processor}

\author{Chao Song$^{1}$}
\author{Jing Cui$^{1}$}
\author{H. Wang$^{1}$}
\author{J. Hao$^{2}$}
\author{H. Feng$^{2}$}
\author{Ying Li$^{3}$}
\email{yli@gscaep.ac.cn}
\affiliation{$^1$ Interdisciplinary Center for Quantum Information and Zhejiang Province Key Laboratory of Quantum Technology and Device, Department of Physics, Zhejiang University, Hangzhou 310027, China, \\
$^2$ Institute of Automation, Chinese Academy of Sciences, Beijing 100190, China, \\
$^3$ Graduate School of China Academy of Engineering Physics, Beijing 100193, China
}
\date{\today}

\begin{abstract}
Medium-scale quantum devices that integrate about hundreds of physical qubits are likely to be developed in the near future. However, such devices will lack the resources for realizing quantum fault tolerance. Therefore, the main challenge of exploring the advantage of quantum computation is to minimize the impact of device and control imperfections without encoding. Quantum error mitigation is a solution satisfying the requirement. Here, we demonstrate an error mitigation protocol based on gate set tomography and quasiprobability decomposition. One- and two-qubit circuits are tested on a superconducting device, and computation errors are successfully suppressed. Because this protocol is universal for digital quantum computers and algorithms computing expected values, our results suggest that error mitigation can be an essential component of near-future quantum computation.
\end{abstract}

% \pacs{03.67.Ac}


\maketitle

\section{Introduction}

\begin{figure*}
\centering
\includegraphics[width=1\linewidth]{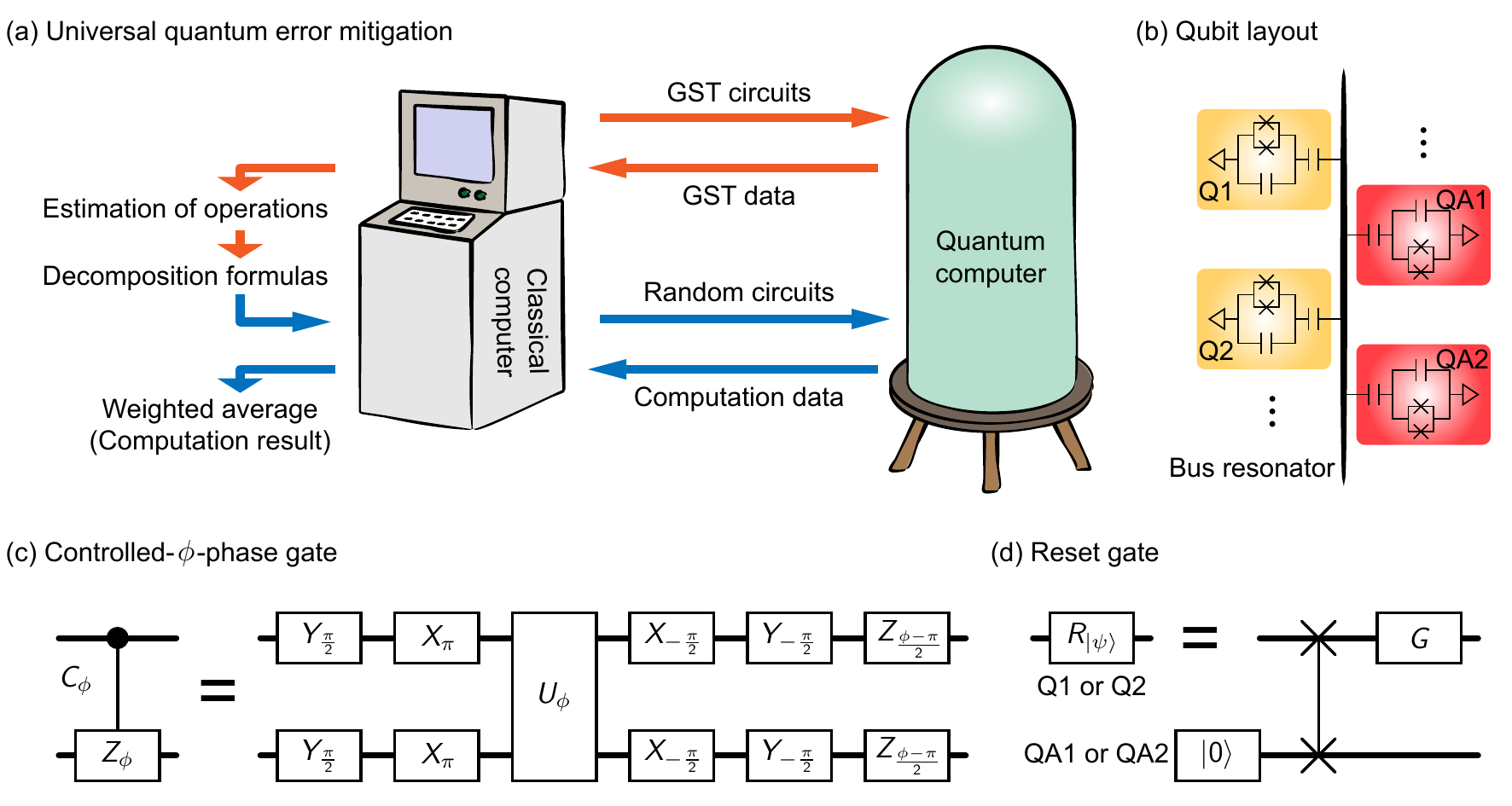}\\
\caption{
(a) Flowchart of the universal quantum error mitigation, which has two-stages: gate set tomography (GST) and the random circuit computation. (b) Layout of four qubits actively used in the experiment. The information is encoded in Q1 and Q2, and the other two qubits QA1 and QA2 are ancillary qubits. (c) Controlled-$\phi$-phase gate $C_\phi$ realized using the dressed state gate $U_{\phi}$ and single-qubit gates. The single-qubit gate $P_\theta = e^{-i\frac{\theta}{2}P}$, where $P =X,Y,Z$. (d) Reset gate $R_{\ket{\psi}}$ realized using an ancillary qubit QA1 or QA2, which reinitialize the qubit Q1 or Q2 in the state $\ket{\psi} = G\ket{0}$.
}
\label{figure1}
\end{figure*}

Quantum computers are quantum-mechanical devices capable of solving problems that are believed to be intractable for classical computers. The most essential issue in practicing quantum computation is to deal with imperfections of the device and control that cause computation errors. Quantum error correction can suppress the chance of errors to an arbitrarily low level, which however is beyond the scope of near-future technologies~\cite{Fowler2012, Joe2016}. For shallow algorithms executed on near-future quantum devices~\cite{Peruzzo2014, OMalley2016, Shen2017, Kandala2017, Colless2018, Hempel2018, Kandala2018}, quantum error mitigation (QEM) methods~\cite{Li2017, Temme2017} are recently proposed to attain a computation result with error minimized, which are practical because encoding is not required. The QEM protocol based on a combination of gate set tomography and quasiprobability decomposition is one of those methods~\cite{Endo2018}, which can be applied to any platform without prior knowledge of imperfections and works for any algorithm that outcomes are expected values of certain observables. In this method, certain quantum circuits are first executed to identify a model of imperfections, then random circuits are sampled from a distribution according to the model [see Fig.~\ref{figure1}(a)]. The theory suggests that the average of such random circuits can provide an accurate computation result. Here, we demonstrate the experimental realization of this method for the first time. The device is a superconducting circuit consisting of ten frequency-tunable transmon qubits, among which four qubits are actively used in the demonstration. Details of the device can be seen in references~\cite{Song2017}. We use single- and two-qubit circuits to test this method and find that with error mitigation the computation accuracy is significantly improved.

We utilize gate set tomography (GST)~\cite{Merkel2013, Greenbaum2015, BlumeKohout2017} to acquire information about the measurement and gate errors in the experiment, which is then used in QEM to decompose any ideal measurement or gate by those experimentally accessible ones with errors. GST can be seen as a self-consistent extension of the quantum process tomography, which takes into account all the errors occurred in the experimental operations including state preparations, quantum gates and measurements.

We use Pauli transfer matrix (PTM) representation to notate quantum states, quantum gates and measurements as commonly adopted in quantum tomography. We define $\sigma_i$ as the $i^{\rm th}$ operator from the $n$-qubit Pauli basis $\mathcal{P}=\{I, X, Y, Z\}^{\otimes n}$, where $I, X, Y, Z$ denote the identity and three Pauli matrices, respectively. In PTM representation, a state $\rho$ is expressed as a column vector $|\rho\rangle\rangle$ with elements $|\rho\rangle\rangle_i={\rm Tr}(\sigma_i\rho)$; An observable $Q$ is expressed as a row vector $\langle\langle Q|$ with elements $\langle\langle Q|_i={\rm Tr}(\sigma_i Q)/2^n$; The superoperator $\mathcal{U}$ of a gate is expressed as a matrix $\mathcal{U}_{ij}={\rm Tr}[\sigma_i\mathcal{U}\sigma_j]/2^n$. Thus the expected value of the observable $Q$ in the state $\rho$ going through a sequence of gates $\mathcal{U}_1,...,\mathcal{U}_N$ reads as follows:
$${\rm Tr}[Q\mathcal{U}_N...\mathcal{U}_1 \rho] = \langle\langle Q|\mathcal{U}_N...\mathcal{U}_1|\rho\rangle\rangle.$$

In GST, quantum gates are reconstructed in a set of experiments. In each experiment, one of the gates is applied on an initial state, and then a measurement is performed to read the value of an observable [see Fig.~\ref{figure2}(a)]. We select $\{|0\rangle, |1\rangle, |0+1\rangle, |0-i1\rangle\}^{\otimes n} = \{ \rho_j \}$ as initial states and Pauli operators in $\mathcal{P}$ as observables. Using the PTM representation, we can express the set of initial states as the state preparation matrix $A^{\rm exp}_{i,j}=\langle\langle\sigma_i|\rho^{\rm exp}_j\rangle\rangle$ and similarly express the set of observables as the readout matrix $B^{\rm exp}_{i,j}=\langle\langle\sigma^{\rm exp}_i|\sigma_j\rangle\rangle$. Here and below the operations with no superscripts are ideal and those with the superscript ``exp'' are physical and with errors as experimentally realized or measured. The Gram matrix $g^{\rm exp}=B^{\rm exp}A^{\rm exp}$ and the matrix characterizing the superoperator $\mathcal{U}$, i.e., $\widetilde{\mathcal{U}}^{\rm exp} = B^{\rm exp} \mathcal{U}^{\rm exp} A^{\rm exp}$, can be obtained by applying the relevant operations in sequence in the experiment.

Experimental operations of $A^{\rm exp}$, $B^{\rm exp}$ and $\mathcal{U}^{\rm exp}$ can be reconstructed by analysing the data of $g^{\rm exp}$ and $\widetilde{\mathcal{U}}^{\rm exp}$. However, we cannot exactly reconstruct experimental operations due to the insufficient information encoded in $g^{\rm exp}$ and $\widetilde{\mathcal{U}}^{\rm exp}$ in the presence of both state preparation and measurement errors. In our device, fidelities of the state preparation and single-qubit gates are much higher than other operations. Therefore, we assume that initial states are error-free and take $\widehat{A} = \langle\langle\sigma_i|\rho_j\rangle\rangle$ as a decent guess of $A^{\rm exp}$, where the caret symbol is introduced to differentiate an estimate from the physical operation itself. Then according to the linear inversion method~\cite{Merkel2013, Greenbaum2015, BlumeKohout2017}, we have the estimates for $B^{\rm exp}$ and $\mathcal{U}^{\rm exp}$ as $\widehat{B} = g^{\rm exp}\widehat{A}^{-1}$ and $\widehat{\mathcal{U}} = \widehat{B}^{-1}\widetilde{\mathcal{U}}^{\rm exp}\widehat{A}^{-1}$, respectively. The difference between the physical gate $\mathcal{U}^{\rm exp}$ and its estimate $\widehat{\mathcal{U}}$ depends on state preparation errors (i.e.~the diffrence between $A^{\rm exp}$ and $\widehat{A}$). We note that this difference is not important because of not only the high fidelity of state preparation in our device but also the self-consistency of GST. Even if state preparation errors are significant, assuming the error-free state preparation does not affect the accuracy of quantum computation using QEM~\cite{Endo2018}. With these results obtained from the GST experiment, we can decompose any ideal measurement and gates into experimentally achieved operations.

Given the decomposition formulas~\cite{Endo2018}, we randomly generate circuits modified from the original circuit of the computation task and implement these random circuits to obtain the computation result with error mitigated, because errors in random circuits cancel with each other when taking the average. In this paper, we only decompose and replace the measurement and two-qubit gate in the original circuit while the state preparation and single-qubit gates are unchanged. We do this because the error of the previous two is larger than the latter by an order of magnitude. To be explicit, by heralding the ground state for qubit preparation and measuring the heating rate~\cite{Chen2016}, we estimate the ground state preparation error to be below $0.25\%$; error rates of single-qubit gates are calibrated to be below $0.25\%$ in randomized benchmarking~\cite{Emerson2005, Knill2008, Barends2014}. In comparison, the readout error is about $3.5\%$ for the ground state and $5.7\%$ for the excited state; the two-qubit gates applied in this paper have errors around $7\%$. We remark that gate fidelities can be further boosted on our device, however, which is unnecessary for the purpose of QEM demonstration.

%\section{Results}

\section{Mitigating readout error in one-qubit computation}

\begin{figure}
\centering
\includegraphics[width=1\linewidth]{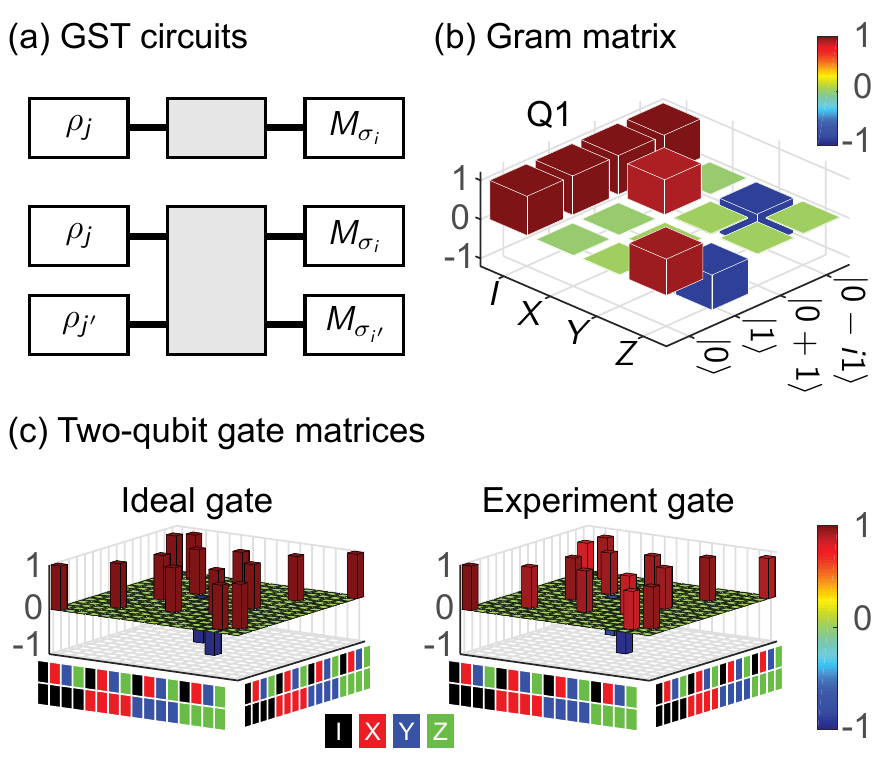}
\caption{
(a) One- and two-qubit gate set tomography (GST) circuits. The gate to be characterized (marked in gray) is implemented in between the state preparation and measurement. Gram matrices and matrices of measurement-initialization gates are obtained using the one-qubit circuit. Matrices of two-qubit gates are obtained using the two-qubit circuit. For the gram matrix, the gate is null. (b) Gram matrix $g^{\rm exp}$ of the qubit Q1. (c) Pauli transfer matrices of the two-qubit gate $C_\pi$. For the ideal gate, each element is calculated as $\mathcal{U}_{ij} = {\rm Tr}[\sigma_{i}\mathcal{U}\sigma_{j}]/4$, where $\mathcal{U}$ is the ideal superoperator for $C_\pi$. For the experiment gate, the matrix is $\widehat{\mathcal{U}}$ the result of GST.
}
\label{figure2}
\end{figure}

We first test the effect of QEM with a one-qubit computation, whose circuit is shown in Fig.~\ref{figure3}(a). In this circuit, we initialize the qubit by heralding the state $|0\rangle$, with a state fidelity above $0.997$. Then the gate $X_{\pi/2} = e^{-i\frac{\pi}{4}X}$, whose gate fidelity is calibrated to be $0.998$ by randomized benchmarking, is applied to rotate the qubit state around the x-axis by $\pi/2$, after which the operator $Z$ is measured. The measurement is denoted by $M_Z$ in the circuit, which yields an outcome of $-1$ or $1$. We repetitively implement the circuit for $3,000$ times and take the average as the expected value of $Z$. We then repeat the same procedure to obtain $100$ expected values. The histogram analysis of the resulting data is shown in the upper panel of Fig.~\ref{figure3}(f). The average over $100$ expected values gives $\langle Z \rangle^{\rm exp} = 0.027 \pm 0.016$.

Since the state preparation and single-qubit gate are both quite precise, the relatively large deviation of $\langle Z \rangle^{\rm exp}$ from zero, i.e.~the ideal result, is mainly due to the readout error, for which we intend to mitigate by decomposing the readout operation. To work out the decomposition formula, we obtain the Gram matrix $g^{\rm exp}$ in the experiment, and the result is shown in Fig.~\ref{figure2}(b). Assuming the error-free state preparation, we can take a reasonable estimate of $A^{\rm exp}$ as
\begin{equation*}
\widehat{A}=\left(
\begin{array}{cccc}
1 & 1 & 1 & 1\\
0 & 0 & 1& 0\\
0 & 0 & 0 & -1\\
1 & -1 & 0& 0\\
\end{array}
\right),
\end{equation*}
where the Pauli operator basis is sorted in the order $I,X,Y,Z$. With $\widehat{B} = g^{\rm exp}\widehat{A}^{-1}$, we can decompose the observable $Z$ in the form $Z = \sum_i q_i \widehat{\sigma}_i$, where $\langle\langle\widehat{\sigma}_i|$ are rows of $\widehat{B}$, and the quasi-probabilities $q_i$ are elements of the vector $q = \langle\langle Z|\widehat{B}^{-1}$. Quasi-probabilities are real but can be greater than unity and be negative. In the quantum computation with QEM, we implement random circuits for $3,000$ times. In each of them, the measurement of $Z$ in the original circuit is replaced by the measurement of $\sigma_i^{\rm exp}$ [see Fig.~\ref{figure3}(b)] with the probability $|q_i|/\sum_l|q_l|$. Here, $\sigma_i^{\rm exp}$ is the physical measurement whose estimate is $\widehat{\sigma}_i$. The measurement outcome is $+1$ or $-1$ (the outcome is always $+1$ in the measurement of $\sigma_0^{\rm exp}$, i.e., the identity operator $I$). When taking the average of measurement outcomes, each outcome is multiplied by a weight factor of $w_i = {\rm sgn}(q_i)\cdot\sum_l|q_l|$. We take this weighted average as the expected value of $Z$. Then we repeat the above procedure to obtain $100$ expected values, and the average over all these expected values yields $\langle Z \rangle^{\rm QEM} = 0.003 \pm 0.022$ as shown in the lower panel of Fig.~\ref{figure3}(f). The accuracy of the computation is successfully improved compared with the computation without QEM.

\section{Mitigating readout and entangling gate errors in two-qubit computation}

\begin{figure*}
\centering
\includegraphics[width=1\linewidth]{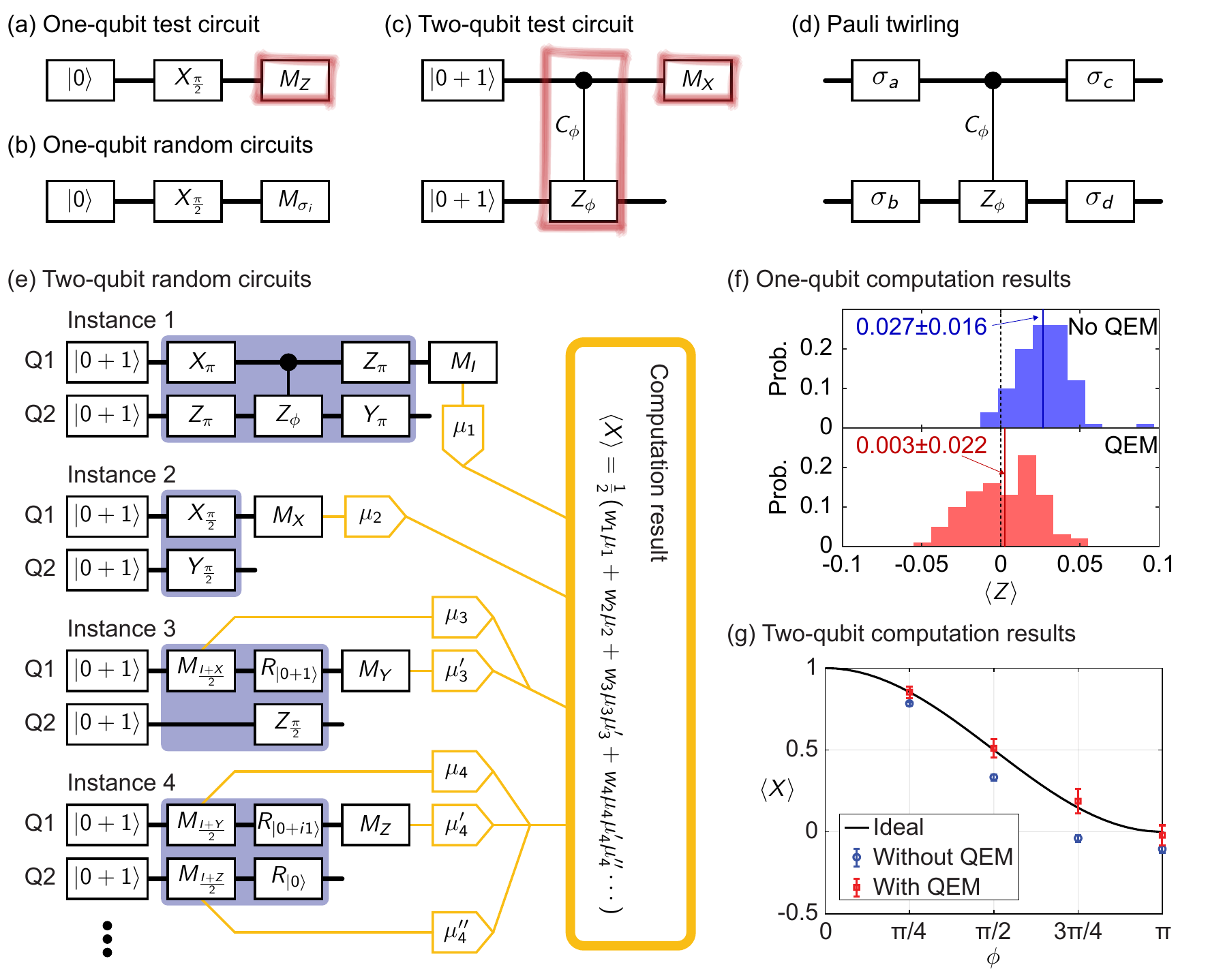}\\
\caption{
(a) Circuit of the one-qubit computation. In the quantum error mitigation (QEM), the measurement of the observable $Z$ is replaced by random gates. (b) Random circuit of the one-qubit computation, in which the measurement in the original circuit is replaced by the measurement of the observable $\sigma_i$. (c) Circuit of the two-qubit DQCp computation. In QEM, the two-qubit gate and the measurement are replaced by random operations. (d) Circuit of the Pauli twirling. (e) Representative Random circuits of the two-qubit computation. $\mu$ denotes the outcome of the corresponding measurement, and $w$ is the weight of the corresponding circuit $w_{j,i}$ as defined in the main text (in the figure the subscript of $w$ denotes the number of the instance). The circuit in the blue box is the replacement of the two-qubit gate $C_\phi$. We note that in the instance 1 four single-qubit gates are Pauli gates of the Pauli twirling. (f) Results of the one-qubit computation. The probability distribution of the computation result is plotted. Without error, the ideal result is $\langle Z \rangle = 0$. (g) Results of the two-qubit computation. Each data point is obtained using $1,000,000$ instances. We implement random circuits for $10,000$ times to compute one average value of $X$ and repeat the computation to obtain $100$ average values. The error bar indicates the standard deviation of these average values.
}
\label{figure3}
\end{figure*}

Now we turn to a two-qubit computation, taking the deterministic quantum computation with pure states (DQCp)~\cite{Knill1998} as an example. The circuit is shown in Fig.~\ref{figure3}(c). A main error source in this circuit is the controlled-$\phi$-phase gate $C_\phi = \frac{I+Z}{2}\otimes I + \frac{I-Z}{2}\otimes Z_\phi$, where $Z_\phi = e^{-i\frac{\phi}{2}Z}$. This gate is realized with a two-qubit dressed state gate $U_{\phi}$~\cite{Guo2018} plus ten single-qubit gates as shown in Fig.~\ref{figure1}(c). The two-qubit dressed state gate essentially achieves a controlled-$\phi$-phase gate in the $X$ basis, and single-qubit gates are used to transform the $X$ basis into the $Z$ basis. Fidelities of different $C_\phi$ gates are $0.958\pm0.010$, $0.935\pm0.011$, $0.920\pm0.011$, and $0.915\pm0.011$ for $\phi = \pi/4$, $\pi/2$, $3\pi/4$, and $\pi$, respectively, characterized using GST. We remark that gate fidelities can be further boosted on our device, however, which is unnecessary for the purpose of QEM demonstration.

To mitigate the error in $C_\phi$, we need first to work out the decomposition formula. Given the ideal superoperator $\mathcal{U}$ representing $C_\phi$, the decomposition formula reads $\mathcal{U} = \sum_j q'_j \widehat{\mathcal{U}}_j$, where $q'_j$ is a set of quasi-probabilities as introduced previously and $\widehat{\mathcal{U}}_j$ is the estimate of a set of experimentally achieved operations $\mathcal{U}^{\rm exp}_j$ as defined below and in Table.~\ref{table}.

\begin{table}
\caption{
16 single-qubit basis operations. $P_{\theta} = e^{-i\frac{\theta}{2}P}$ denotes the gate of rotation along the $P$-axis by an angle of $\theta$, where $P=X,Y,Z$. $M_P$ denotes the operation of measuring the eigenvalue of the Pauli operator $P$ whose outcomes are $\pm 1$. $M_{\frac{I+P}{2}}$ and $M_P$ are the same operation but outcomes are noted differently, and $M_{\frac{I+P}{2}}$ denotes the operation of measuring the eigenvalue of the operator $\frac{I+P}{2}$ whose outcomes are $0$ and $1$. $R_{|\psi\rangle}$ denotes the operation of resetting the qubit state to $|\psi\rangle$. For composed operations, operations are implemented from left to right in sequence.
}
\begin{ruledtabular}
\begin{tabular}{ c c c c c c }
 No. & Operation & No. & Operation \\
 \hline
 1 & $I$                            & 9& $X_{\pi}, Y_{-\frac{\pi}{2}}$ \\
 2 & $X_{\pi}$                & 10& $Y_{\pi}, X_{\frac{\pi}{2}}$ \\
 3 & $Y_{\pi}$                & 11& $M_{\frac{I+X}{2}}, R_{|0+1\rangle}$ \\
 4 & $Z_{\pi}$                & 12& $M_{\frac{I+X}{2}}, R_{|0-1\rangle}$ \\
 5 & $X_{\frac{\pi}{2}}$            & 13& $M_{\frac{I+Y}{2}}, R_{|0+i1\rangle}$ \\
 6 & $Y_{\frac{\pi}{2}}$            & 14& $M_{\frac{I+Y}{2}}, R_{|0-i1\rangle}$\\
 7 & $Z_{\frac{\pi}{2}}$                & 15& $M_{\frac{I+Z}{2}}, R_{|0\rangle}$ \\
 8 & $X_{\pi}, Z_{\frac{\pi}{2}}$    & 16& $M_{\frac{I+Z}{2}}, R_{|1\rangle}$ \\
\end{tabular}
\end{ruledtabular}
\label{table}
\end{table}

In our experiment, $257$ operations are used for decomposing an ideal $C_\phi$ gate. The first $256$ operations are generated from the tensor product of $16$ single-qubit operations, which include measurement and reset gates, as listed in Table.~\ref{table}. The $257^{\rm th}$ operation is the gate $C_\phi$ modified by the Pauli twirling as we will explain soon. We reconstruct the experimental operations of $C_\phi$ and single-qubit measurement-reset gates in GST, while we simply assume single-qubit gates are error-free, because single-qubit gates can be experimentally implemented with high fidelity. Unlike the state preparation, assuming error-free single-qubit gates can potentially cause inaccuracy in the quantum computation with QEM. PTM of the controlled-$\pi$-phase gate $C_\pi$ obtained using GST is illustrated in Fig.~\ref{figure2}(c) as an example. These $257$ operations are linearly independent, which ensures that the decomposition solutions can always be found by solving a system of linear equations. In all solutions we choose the one with the minimum in $\sum_k|q'_k|$ to minimize the variance of the computation result.

The measurement-reset operation is occasionally in random circuits [see Table.~\ref{table}]. In order to minimize the time of reset, we realize the reset gate using ancilla qubits QA1 and QA2 on the same chip [see Fig.~\ref{figure1}(b)]. Each reset operation uses an ancilla qubit initially prepared in the ground state $|0\rangle$, then a swap gate is applied to reinitialize the target qubit when the reset is requested~\cite{Song2017}, following which a single-qubit gate $G$ with the fidelity above $0.997$ rotates the qubit to the state $\ket{\psi}$ as shown in Fig.~\ref{figure1}(d). The whole measurement-reset operation typically has a fidelity of around $0.916$.

Pauli twirling converts the error in a gate into \emph{stochastic} Pauli errors~\cite{Knill2004, Wallman2016, OGorman2016}, which can reduce the variance of the computation result~\cite{Endo2018}. The circuit of the gate $C_\phi$ with Pauli twirling is shown in Fig.~\ref{figure3}(d). We sandwich $C_\phi$ in between four Pauli gates (two for each qubit) which are randomly chosen but conserves the gate $C_\phi$ up to a global phase difference. If all gates are error-free, the two-qubit gate realized in this way is still $C_\phi$, i.e.~$[C_\phi] = \sum_{a,b} p_{a,b} [\sigma_c\otimes\sigma_d C_\phi \sigma_a\otimes\sigma_b]$. Here we use the bracket notation to denote a superoperator $[C](\rho) = C \rho C^\dag$, $\{ \sigma_i \}$ are single-qubit Pauli gates chosen to satisfy $\sigma_c\otimes\sigma_d C_\phi \sigma_a\otimes\sigma_b = \eta C_\phi$, $\eta$ can be any phase factor, and $p_{a,b}$ is the probability. If gates have errors, the two-qubit gate will be effectively changed by the Pauli twirling. Using the twirled gate as the $257^{\rm th}$ operation in the decomposition, we need the estimate of the twirled gate, which is $\widehat{\mathcal{U}}' = \sum_{a,b} p_{a,b} [\sigma_c\otimes\sigma_d] \widehat{\mathcal{U}} [\sigma_a\otimes\sigma_b]$, where $\widehat{\mathcal{U}}$ is the estimate of the experimental operation of $C_\phi$ obtained in GST. For $C_\pi$, the distribution of Pauli gates is unifrom, i.e.~$p_{a,b} = 1/16$. For other $C_\phi$ gates where $\phi \neq \pi$, Pauli gates are chosen from a subset: We take $p_{a,b} = 1/4$ if $\sigma_a\otimes\sigma_b \in \{I, Z\}^{\otimes 2}$, and $p_{a,b} = 0$ otherwise.

In the two-qubit computation with QEM, to mitigate both readout and two-qubit gate errors, the gate $C_\phi$ is randomly replaced by the gate $\mathcal{U}^{\rm exp}_j$ with the probability $|q'_j|/\sum_k|q'_k|$, and the measurement of $X$ is replaced by the measurement of $\sigma_i^{\rm exp}$ with the probability $|q''_i|/\sum_l|q''_l|$, where $q'' = \langle\langle X|\widehat{B}^{-1}$. Similar to the one-qubit case, the measurement outcome of each random circuit is multiplied by a weight factor of $w_{j,i} = {\rm sgn}(q'_j q''_i)\cdot\sum_k|q'_k| \cdot\sum_l|q''_l|$, and we take the weighted average as the computation result as shown in Fig.~\ref{figure3}(e).

In the experiment, we adjust the phase $\phi$ of $C_\phi$ and measure the expected value of $X$. When implementing the computation with QEM, we randomly sample a circuit according to the decompositions of both $C_\phi$ and the observable $X$. Representative decomposed sampling circuits are shown in Fig.~\ref{figure3}(e). The experiment result is shown in Fig.~\ref{figure3}(g), which demonstrates a significant improvement on the computation accuracy.

The most significant improvement is obtained at $\phi = \pi/2$, in which case the difference between the computation result and the ideal value is reduced from $0.1690$ to $0.0102$ by using QEM. To estimate the fidelity required to achieve the same computation accuracy, we consider a quantum system with depolarizing error channels~\cite{Knill2005} and assume that the state preparation and single-qubit gates are ideal. The depolarizing error channel either preserves or completely destroys the information with certain probabilities~\cite{Supp}, which does not characterize our device. We choose the depolarizing model because it takes all possible errors into account with equal probability. In the depolarizing model, the two-qubit gate and measurement with the fidelity $\sim 99.3\%$ are required to achieve the computation accuracy $0.0102$, which is comparable to the highest fidelity reported in the superconducting qubit system~\cite{Barends2014, Jeffrey2014}.

\section{Discussions}

For multi-qubit devices, GST of the entire device is not practical, because the experiment time increases exponentially with the qubit number. Similarly, the number of operations for decomposing a multi-qubit gate also increases exponentially with the qubit number. Single-qubit and two-qubit gates are sufficient for the universal quantum computation. Therefore, if errors are uncorrelated, we only need to implement GST up to two qubits and decompose two-qubit gates, as demonstrated in our experiment. Errors are uncorrelated if the evolution of two qubits under a two-qubit gate is independent from the evolution of other qubits. As a result of the evolution, the quantum operation on the entire device can be factorized into the product of an operation on the two qubits and operations on other qubits. It is similar for single-qubit gates. In our device, the primary dephasing noise is dominated by fluctuators in the form of magnetic moments, whose influence is local in each individual physical qubit, and, therefore, the dephasing-induced errors are uncorrelated between qubits. In our experiment, we have neglected error correlations in GST, so that single-qubit operations are characterized in single-qubit tomography even in the two-qubit experiment. Neglecting error correlations sacrifices accuracy of QEM. A significant effect of correlations on the computation result is not observed in our experiment.

We have experimentally demonstrated that the universal QEM protocol can significantly reduce the error in quantum computation on a noisy quantum device. The protocol in our experiment does not require sub-threshold error rate or tremendous additional physical qubit resource. Compared with the algorithm-specified protocol~\cite{Colless2018} and the extrapolation of gate time~\cite{Kandala2018}, the combination of gate set tomography and quasi-probability decomposition is not restricted to the algorithm or error model. A few techniques in QEM are explored: estimate of the state preparation matrix according to ideal states, Pauli twirling for randomizing the error, approximate GST and decomposition neglecting error correlations. An important factor limiting the circuit depth in our demonstration is the variance of computation result, which depends on the error rate of quantum gates. Improvement in gate fidelity can extend the circuit depth, and relatively deep circuits can be implemented on intermediate-scale devices with a feasible fidelity~\cite{Endo2018}. By demonstrating the power of error mitigation techniques on the superconducting quantum device, our results highlight the potential of using such techniques in computation tasks on near-future quantum devices.

\noindent{\textbf{Acknowledgments}}\\
\noindent{\footnotesize{
We thank Xiaobo Zhu and Dongning Zheng for fabricating the device. This work was supported by the National Basic Research Program of China (Grant No. 2017YFA0304300) and the National Natural Science Foundations of China (Grants No. 11434008 and No. 11725419). YL is supported by NSAF (Grant No. U1730449).
}}

%%%%%%%%%% Merge with supplemental materials%%%%%%%%%%%
\pagebreak

\begin{center}
\textbf{\large Supplementary Information for "Quantum computation with universal error mitigation on superconducting quantum processor"}
\end{center}

\setcounter{figure}{0}
\setcounter{table}{0}
\renewcommand\thefigure{S\arabic{figure}}
\renewcommand\thetable{S\arabic{table}}

\section{Randomized benchmarking for single-qubit gates}

\begin{figure}
  \centering
  \includegraphics[width=1\linewidth]{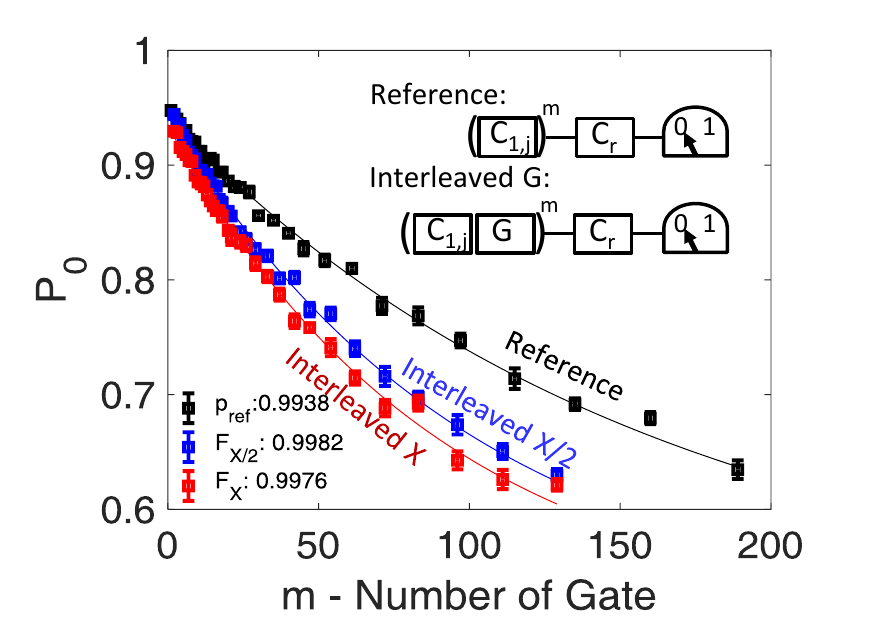}\\
  \caption{Randomized benchmarking data. We fit the data with $P_0\propto p_{ref,G}^m$, and the gate fidelity is calculated as $F=1-0.5(1-p_{G}/p_{ref})$. $G$ denotes the gate to be characterized. The {\it Reference} and interleaved sequences are shown in the inset, where $C_{1,j}$ is the $j^{\rm th}$ randomly selected gate from the single-qubit Clifford group, and $C_r$ is the recovery gate from the same group bringing the qubit back to the state $|0\rangle$.}
  \label{figure1_supp}
\end{figure}

In Fig.~\ref{figure1_supp} we show the randomized benchmarking data for gates $X_{\pi}$ and $X_{\frac{\pi}{2}}$ of Q1. We find that fidelities of these two gates are $0.9982$ and $0.9976$, respectively.

\section{Heating rate measurement}

\begin{figure}
  \centering
  \includegraphics[width=1\linewidth]{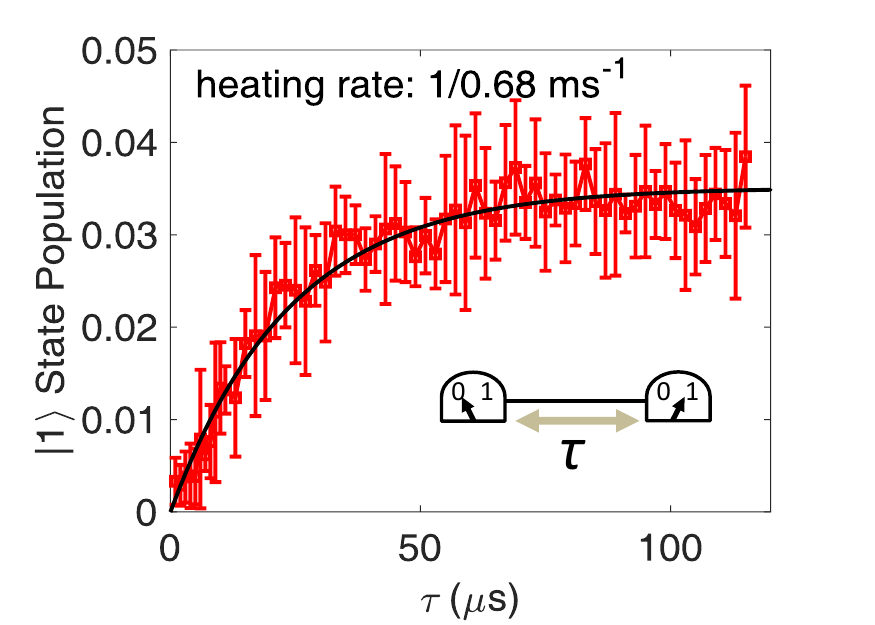}\\
  \caption{Heating rate measurement. The sequence for measuring heating rate is shown in the inset. The first measurement herald the ground state. After the delay $\tau$, we measure the probability of the qubit being in the excited state. The black line is a fit to the data using the formula $P_1 = \frac{\Gamma_{\uparrow}}{\Gamma_{\uparrow}+\Gamma_{\downarrow}}\left[ 1-e^{-(\Gamma_{\uparrow}+\Gamma_{\downarrow})\tau} \right]$, where $P_1$ is the $|1\rangle$ state population after subtracting readout error and $\Gamma_{\uparrow}, \Gamma_{\downarrow}$ are heating and decay rates of the qubit. We take $\Gamma_{\downarrow} = 1/T_1$ and optimize $\Gamma_{\uparrow}$ to fit the data. In our experiment, we wait $1.6\text{ }\mu{\rm s}$ after heralding for the readout resonator to settle down. Taking $ T_1 = 24.7 \mu s$ and $\tau=1.6\text{ }\mu{\rm s}$ in the equation, we obtain the ground state preparation error of $0.0023$.}
  \label{figure2_supp}
\end{figure}

In the experiment we herald the state $|0\rangle$ to initialize the qubit. We need to wait $1.6\text{ }\mu{\rm s}$ after the measurement pulse for the readout resonator to settle down before any further operation can be applied, during which the qubit may be heated again, causing the ground state preparation error. In Fig.~\ref{figure2_supp}, we show the measurement of the heating rate of Q1, which indicates a ground state preparation error below $0.0025$.

\section{Readout error for Q1 and Q2.}

In order to measure the readout error, we repeatedly prepare the qubit to the state $|0\rangle$ or $|1\rangle$ and measure the probability of incorrect output. The results are shown in Table.~\ref{table1_supp}

\begin{table}
 \caption{Error rates of readout measured by repeatedly preparing the state $|0\rangle$ or $|1\rangle$ and measuring the probability of incorrect output.}
 \begin{ruledtabular}
 \begin{tabular}{ c c c }
  & Q1 & Q2 \\
 \hline
    error rate of ground state readout      & 0.0343 & 0.0360  \\
    error rate of excited state readout      & 0.0526 & 0.0607  \\
 \end{tabular}
 \end{ruledtabular}
 \label{table1_supp}
\end{table}

\section{One-qubit QEM experiment}

\begin{figure}
  \centering
  \includegraphics[width=1\linewidth]{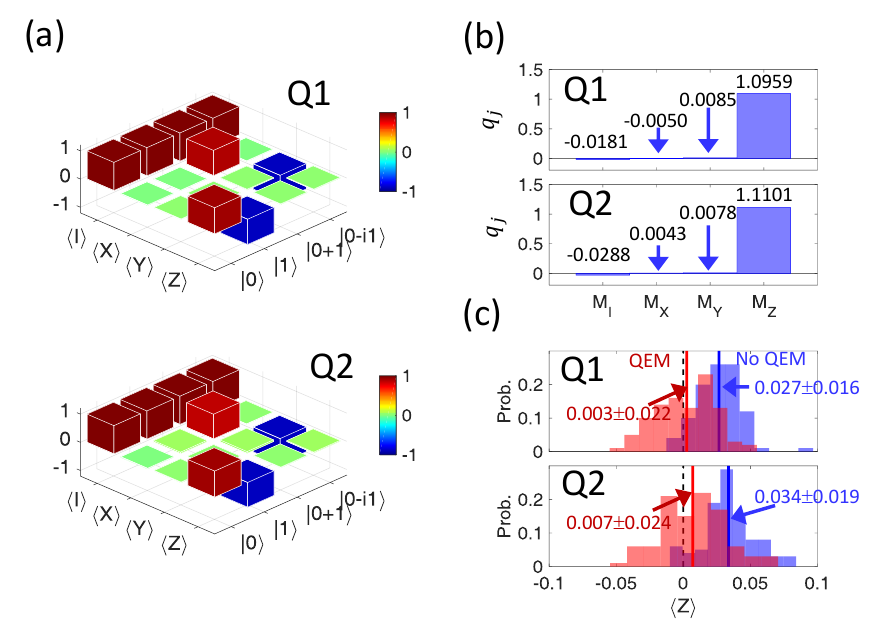}\\
  \caption{One-qubit QEM experiment. (a) Gram matrices for Q1 and Q2. (c) Quasiprobability decomposition coefficients. (d) One-qubit QEM experiment results for Q1 and Q2.}
  \label{figure3_supp}
\end{figure}

Gram matrices $g$ for Q1 and Q2 are shown in Fig.~\ref{figure3_supp}(a). The decomposition coefficients (i.e.~quasiprobabilities) of the measurement $M_Z$ is shown in Fig.~\ref{figure3_supp}(b). In the QEM experiment, we take a sample number of $3,000$ to obtain an average value $\langle Z \rangle$ and repeat the experiment for $100$ times to obtain the distribution of $\langle Z \rangle$. The result is shown in Fig.~\ref{figure3_supp}(c).

\section{Measurement and reset gates}

\begin{figure}
  \centering
  \includegraphics[width=1\linewidth]{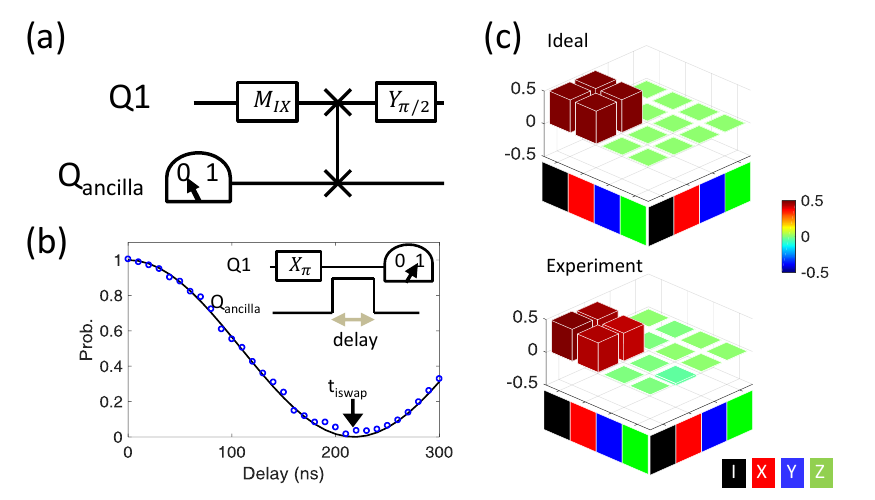}\\
  \caption{(a) Circuit for the measurement-reset gate of $M_{\frac{I+X}{2}}$ and $R_{|0+1\rangle}$. (b) The key element of the reset gate is an i-swap gate, which is achieved by tuning the ancillary qubit, initialized in ground state, on resonance with the target qubit for a while. Because of the virtual coupling through the bus resonator, states of two qubit will be exchanged. The data show the probability in the target state used to determine the length of i-swap gate. The black line is the fitting curve. (c) GST result for the measurement-reset gate on Q1. We find that the gate fidelity is $0.912\pm0.014$. Gate fidelities for all measurement-reset gates are listed in Table.~\ref{table2_supp}}
  \label{figure4_supp}
\end{figure}

In the two-qubit QEM experiment, we used two ancillary qubits QA1 and QA2 to reset Q1 and Q2. The Circuit for the measurement-reset gate of $M_{\frac{I+X}{2}}$ and $R_{|0+1\rangle}$ on Q1 is shown in Fig.~\ref{figure4_supp}(a). An i-swap gate is used in the reset gate, which is shown in Fig.~\ref{figure4_supp}(b). The GST result for this measurement-reset gate is shown in Fig.~\ref{figure4_supp}(c). Gate fidelities for all measurement-reset gates used in this paper are listed in Table.~\ref{table2_supp}.

\begin{table}
 \caption{Gate fidelities for all measurement-reset gates used in this paper. Gates were characterized by using GST. The fidelity was acquired by firstly transforming the Pauli transfer matrix to the $\chi$ matrix and then calculating $F = {\rm Tr}(\chi^{\rm exp} \chi^{\rm ideal})/[{\rm Tr}(\chi^{\rm exp}) {\rm Tr}(\chi^{\rm ideal})]$.}
 \begin{ruledtabular}
 \begin{tabular}{ c c c }
 gate & Q1 & Q2 \\
 \hline
    $M_{\frac{I+X}{2}}, R_{|0+1\rangle}$	& $0.937\pm0.034$	& $0.889\pm0.043$ \\
    $M_{\frac{I+X}{2}}, R_{|0-1\rangle}$	& $0.906\pm0.034$	& $0.871\pm0.031$ \\
    $M_{\frac{I+Y}{2}}, R_{|0+i1\rangle}$	& $0.947\pm0.023$	& $0.893\pm0.022$ \\
    $M_{\frac{I+Y}{2}}, R_{|0-i1\rangle}$	& $0.941\pm0.022$	& $0.909\pm0.019$ \\
    $M_{\frac{I+Z}{2}}, R_{|0\rangle}$		& $0.943\pm0.015$	& $0.903\pm0.013$ \\
    $M_{\frac{I+Z}{2}}, R_{|1\rangle}$		& $0.938\pm0.023$	& $0.915\pm0.019$ \\
 \end{tabular}
 \end{ruledtabular}
 \label{table2_supp}
\end{table}

\section{Decomposing $C_\phi$ gate}

\begin{figure}
  \centering
  \includegraphics[width=1\linewidth]{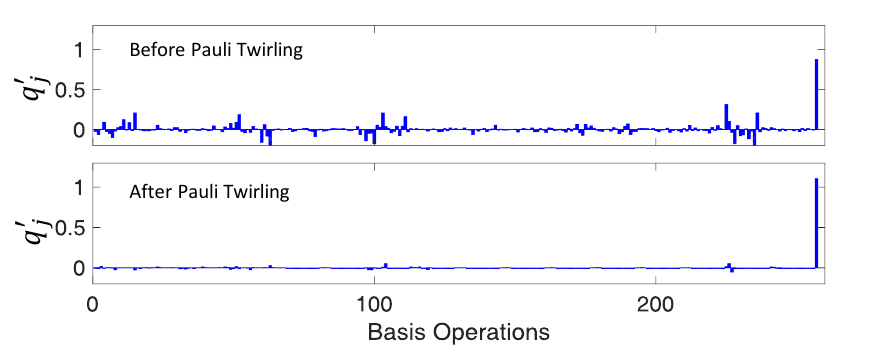}\\
  \caption{Quasiprobability distribution in the decomposition of the $C_\pi$ gate without (up) and with (down) the Pauli twirling.}
  \label{figure5_supp}
\end{figure}

When decomposing two-qubit gates, we suppose that the first $10$ gates listed in Table.~I in the main text are ideal. The last $6$ gates, i.e.~the measurement-reset gates are characterized by using GST, and the GST results are used in the decomposition. Coefficients (i.e.~quasiprobabilities) of the decomposition for the $C_\pi$ gate are shown in Fig.~\ref{figure5_supp}. We find that by applying the Pauli twirling, the cost $\sum_{j}{|q'_j|}$ reduced from $8.71$ to $1.77$. In the two-qubit QEM experiment, we generate random circuits according to the coefficients with the Pauli twirling applied.

\section{Depolarizing error channels}

For a two-qubit gate, if the fidelity is $F_2$ and the error channel is depolarizing, the gate succeeds (i.e.~the gate does not cause any error) with the probability $1-\epsilon_2 = (16F_2-1)/15$ and fails (i.e.~the two-qubit state becomes the maximally mixed state) with the probability $\epsilon_2 = 16(1-F_2)/15$. For a measurement, if the fidelity is $F_{\rm M}$ and the error channel is depolarizing, the measurement succeeds (i.e.~the outcome is true) with the probability $1-\epsilon_{\rm M} = 2F_{\rm M}-1$, and the measurement fails (i.e.~the outcome is completely random) with the probability $\epsilon_{\rm M} = 2(1-F_{\rm M})$. Suppose that $\langle X \rangle^{\rm ideal}$ is the computation result when all operations are error-free, the computation result becomes $\langle X \rangle^{\rm ideal} (1-\epsilon_2)(1-\epsilon_{\rm M})$ when error channels are switched on, and the difference from the ideal value is $\Delta \langle X \rangle \simeq \langle X \rangle^{\rm ideal}(\epsilon_2+\epsilon_{\rm M}) \simeq \langle X \rangle^{\rm ideal}(3-F_2-2F_{\rm M})$. When $\phi = \pi/2$, $\langle X \rangle^{\rm ideal} = 1/2$, therefore $\Delta \langle X \rangle \simeq 0.01$ if $F_2 = F_{\rm M} = 0.993$.

\end{document}